\documentclass[english,twocolumn,amsmath,amssymb,nofootinbib,superscriptaddress,showpacs]{revtex4}

\usepackage[active]{srcltx}

\usepackage{amsfonts}
\usepackage{amssymb}
\usepackage{latexsym}
\usepackage{graphicx}
\usepackage{epsfig,subfigure}
\usepackage{epstopdf}
\usepackage[active]{srcltx}
\makeatother

\begin{document}

\title{Cross-over in non-standard random-matrix spectral fluctuations without unfolding}

\author{G. Torres-Vargas}
\email[]{gamaliel_torres@uaeh.edu.mx}
\affiliation{Instituto de Ciencias B\'asicas e Ingenier\'ia, Universidad Aut\'onoma del Estado de Hidalgo, Pachuca 42184, Hidalgo, Mexico}
\affiliation{Posgrado en Ciencias Naturales e Ingenier\'ia, Universidad Aut\'onoma Metropolitana Cuajimalpa, 05348 CDMX, Mexico}

\author{J. A. M\'endez-Berm\'udez}
\email[]{jmendezb@ifuap.buap.mx}
\affiliation{Instituto de F\'isica, Benem\'erita Universidad Aut\'onoma de Puebla, Apartado Postal J-18, Puebla 72570, Mexico}

\author{J. C. L\'opez-Vieyra}
\email[]{vieyra@nucleares.unam.mx}
\affiliation{Instituto de Ciencias Nucleares, Universidad Nacional Aut\'onoma de M\'exico, 04510 CDMX, Mexico}

\author{R. Fossion}
\email[]{fossion@nucleares.unam.mx}
\affiliation{Instituto de Ciencias Nucleares, Universidad Nacional Aut\'onoma de M\'exico, 04510 CDMX, Mexico}
\affiliation{Centro de Ciencias de la Complejidad (C3), Universidad Nacional Aut\'onoma de M\'exico, 04510 CDMX, Mexico}

\begin{abstract}
Recently, the singular value decomposition (SVD) was applied to standard Gaussian ensembles of Random Matrix Theory (RMT) to determine the scale invariance in the spectral fluctuations without performing any unfolding procedure. Here, SVD is applied directly to the $\nu$-Hermite ensemble and to a sparse matrix ensemble, decomposing the corresponding spectra in trend and fluctuation modes. In correspondence with known results, we obtain that fluctuation modes exhibit a cross-over between soft and rigid behavior. By using the trend modes we performed a data-adaptive unfolding, and we calculate traditional spectral fluctuation measures. Additionally, ensemble-averaged and individual-spectrum averaged statistics are calculated consistently within the same basis of normal modes.
\end{abstract}


\maketitle

\section{Introduction}
\label{SectIntro}

Standard Gaussian ensembles from RMT \cite{meh91} have been enormously successful in the modeling of the fluctuations of quantum excitation spectra \cite{bro81,haa10}, and recently have been used for analyzing the brain functional network \cite{wan16}, as well as in multivariate statistics to model the fluctuations of eigenspectra of correlation matrices of other complex systems \cite{kwa12,mun12,wan15}. More general random-matrix ensembles introduce new statistical features that are absent in standard RMT, such as Gaussian instead of semicircular global eigenvalue densities \cite{flo01}, breakdown of the scale invariance of the long-range fluctuation statistics \cite{jac01,mal07,rel08} and nonergodicity \cite{asa01,jac01,flo01}. Before performing the statistical study of the spectral fluctuations, one has to realize an \emph{unfolding} procedure, which separates the global density of eigenlevels $\overline{\rho}(E)$ from the local fluctuations $\widetilde{\rho}(E)=\rho(E)-\overline{\rho}(E)$. This unfolding procedure is not trivial, and has been shown that the statistical results can be sensitive to the particular unfolding method applied \cite{haa10,jac01,gom02,abu12,abu14,ber16,she18}.

In \cite{fos13}, was proposed a data-adaptive and model-free unfolding procedure, based on SVD, that expresses a spectrum in an exact way as the superposition of global and fluctuation normal modes. When the SVD is applied to standard Gaussian ensembles, the normal modes associated to the fluctuations are scale invariant and obey specific power laws that distinguish between soft and rigid spectra. Moreover, in \cite{tor17}, was showed that applying the SVD method directly to the quantum spectra, it is possible to characterize the transition between the extreme regular and chaotic cases and, hence, quantify the quantum chaos in systems described by RMT in a straightforward way, without implementing any previous unfolding procedure.

In the present contribution, we apply the SVD method directly to a non-standard random-matrix spectra, namely, the $\nu$-Hermite ensemble and a sparse matrix ensemble. We obtain in a direct way that, unlike fluctuations of standard RMT ensembles which are scale invariant and follow a power law \cite{fos13,tor17}, in these cases the scale invariance for the fluctuations is lost, with a cross-over between soft and rigid properties at different scales, in correspondence with known results \cite{jac01,mal07,rel08}, but without implementing any previous standard unfolding procedure, and therefore, avoiding the introduction of possible artifacts. In order to calculate the traditional fluctuation measures, we perform a data-adaptive unfolding of the spectra employing the global modes. In particular, we compare our results for the nearest-neighbor spacing distribution (NNSD) with the distribution of the ratio of two consecutive level spacings $P(\tilde{r})$, which does not depend on the local density of states, and therefore it does not require unfolding \cite{oga07,ata13}. Furthermore, the Fourier power spectrum $P(f)$ of the fluctuations of separate eigenspectra, averaged over the whole ensemble after individual unfolding is applied, is calculated and then compared with the so-called \emph{scree diagram} which is an ensemble-averaged property.

The paper is organized as follows. In Sec. II, we review briefly how the SVD method acts on the ensembles, decomposing each energy spectrum in trend and fluctuation modes. In Sec. III, applying the SVD to the $\nu-$Hermite ensemble, we obtain the scree diagram which shows that the fluctuation modes exhibit a cross-over between soft and rigid behavior. Moreover, we calculate the NNSD, the number variance $\Sigma^2$, the $\Delta_3$ statistics, the distribution $P(\tilde{r})$, as well as the Fourier power spectrum of the fluctuations of separate eigenspectra. In Sec. IV, we apply the SVD to a sparse matrix ensemble, and we also calculate the same fluctuation measures as in the previous section. In Sec. V, we present our conclusions.

\section{Singular Value Decomposition}
\label{Sect_Data-adaptive}

Consider an ensemble of $m=1\ldots M$ level sequences $E^{(m)}(n)$, with $n=1\ldots N$ levels. Each sequence constitutes one of the rows of a $M\times N$ dimensional matrix $\mathbf{X}$, which is interpreted as a multivariate time series,
\begin{equation}
\mathbf{X}=
\left(%
\begin{array}{cccc}
  E^{(1)}(1) & E^{(1)}(2) & \cdots & E^{(1)}(N) \\
  E^{(2)}(1) & E^{(2)}(2) & \cdots & E^{(2)}(N) \\
  \vdots & \vdots & \ddots & \vdots \\
  E^{(M)}(1) & E^{(M)}(2) & \cdots & E^{(M)}(N) \\
\end{array}%
\right).
\label{EqX}
\end{equation}
SVD is a parameter-free matrix decomposition technique that expresses $\mathbf{X}$, in an exact and unique way, as
\begin{equation}
\mathbf{X} = \mathbf{U} \mathbf{\Sigma} \mathbf{V}^T = \sum_{k=1}^r \sigma_k \vec{u}_k \vec{v}_k^T,
\label{EqSVD}
\end{equation}
where $\mathbf{\Sigma}$ is an $M \times N$-dimensional matrix with only diagonal elements which are the ordered \emph{singular values} $\sigma_1 \geq \sigma_2 \geq \ldots \geq \sigma_r$, with $r \leq \mathrm{Min}[M,N]=\mathrm{rank}(\mathbf{X})$. The vectors $\vec{u}_k$ are orthonormal, and they constitute the $k$th columns of the $M \times M$-dimensional matrix $\mathbf{U}$. They are called the \emph{left-singular vectors} of $\mathbf{X}$, and they span its column space. Their physical significance will be explained further on. The vectors $\vec{v}_k$ are orthonormal and they constitute the $k$th columns of the $N \times N$-dimensional matrix $\mathbf{V}$. They are called the \emph{right-singular vectors} of $\mathbf{X}$, and they span its row space, therefore, they constitute a basis of energy \emph{normal modes} for the ensemble. The expression $\vec{u}_k \vec{v}_k^T \equiv \vec{u}_k \otimes \vec{v}_k$ indicates the the outer product of $\vec{u}_k$ and $\vec{v}_k$. A set $\{\sigma_k,\vec{u}_k,\vec{v}_k\}$ is called an \emph{eigentriplet}, and defines completely the eigenmode of order $k$. Any matrix row of $\mathbf{X}$ containing a particular eigenspectrum can be written as,
\begin{equation}
E^{(m)}(n) = \overline{E}(n)+\widetilde{E}(n) = \sum_{k=1}^r \sigma_k U_{mk} \vec{v}^T_k(n),
\label{EqDecomposition}
\end{equation}
where $\lambda_k=\sigma_k^2$, can be interpreted as \emph{partial variances} that indicate how much a specific normal mode $\vec{v}_k$ contributes to the total variance of the ensemble, and the matrix elements $U_{mk}$ serve as coefficients that express a particular level sequence exactly as a weighted sum of normal modes. The normal modes $\vec{v}_k$ with $k=1,\ldots,n_T$, that determine the global spectral properties $\overline{E}$ of a particular spectrum, behave monotonously and can easily be distinguished by their large partial variances $\lambda_k$ that are orders of magnitude larger that the remaining $\lambda_k$ with $k=n_T+1,\ldots,r$ associated to the oscillating normal modes of the fluctuations $\widetilde{E}$. From the log-log plot of the partial variances, known as \emph{scree diagram}, we can see that the oscillating modes follow the power law,
\begin{equation}
\lambda_k\propto1/k^\gamma
\label{EqScree}
\end{equation}
where $\gamma=2$ in the Poisson limit and $\gamma=1$ in the GOE limit \cite{fos13,tor17}.
 
In the following, we apply the SVD method to the cases of the $\nu-$Hermite ensemble and a sparse-matrix ensemble.

\section{$\nu-$Hermite ensemble}
\label{Sect_NuHermite}

The $\nu-$Hermite ensemble or $\nu-$Gaussian ensemble (here we use $\nu$ instead of $\beta$ in order to avoid confusion with the exponent of the power spectrum), is also called the continuous Gaussian ensemble because the parameter $\nu$ interpolates continuously between the classical Gaussian ensembles of RMT, e.g., between the Poisson limit ($\nu=0$) and GOE ($\nu=1$) \cite{rel08,dum02}. One of the most convenient characteristics of the $\nu-$Hermite ensemble is its simple tridiagonal form, which has the important advantage of an unrivaled speedup and efficiency in numerical simulations with large matrix dimensions \cite{mal07,rel08}. A tridiagonal $N \times N$ random matrix from the $\nu-$Hermite ensemble is real and symmetric, and can be defined as,
\begin{widetext}
\begin{equation}
\mathbf{A}_{N,\nu}=\sigma \mathbf{H}_{N,\nu}=\sigma
\left(%
\begin{array}{ccccc}
  H_{11} & H_{12}/\sqrt{2} & 0 & \ldots & 0 \\
  H_{12}/\sqrt{2} & H_{22} & H_{23}/\sqrt{2} & 0 & \ldots \\
  0 & H_{32}/\sqrt{2} & \ldots & \ldots & 0 \\
  \ldots & 0 & \ldots & H_{N-1,N-1} & H_{N-1,N}/\sqrt{2} \\
  0 & \ldots & 0 & H_{N,N-1}/\sqrt{2} & H_{N,N} \\
\end{array}%
\right),
\label{EqNuHermiteMatrix}
\end{equation}
\end{widetext}
where $\sigma$ is a scale factor, and which is chosen as $\sigma=1$ here. The $2N-1$ distinct matrix elements are independent random variables. The $N$ diagonal elements are independently distributed standard normal random variables $\mathcal{N}(0,1)$. The off-diagonal elements $H_{m,m+1} (m=1,\ldots,N-1)$ have a $\chi$ distribution with $m \nu$ degrees of freedom whose probability density is $q_{N,\nu}(x)=2^{1-m \nu/2} x^{m \nu-1} \exp(-x^2/2)/ \Gamma(m \nu /2) (x \geq 0)$.

\begin{figure}[h]
\begin{center}
\begin{minipage}{0.9\linewidth}
\centering
\includegraphics[width=\linewidth]{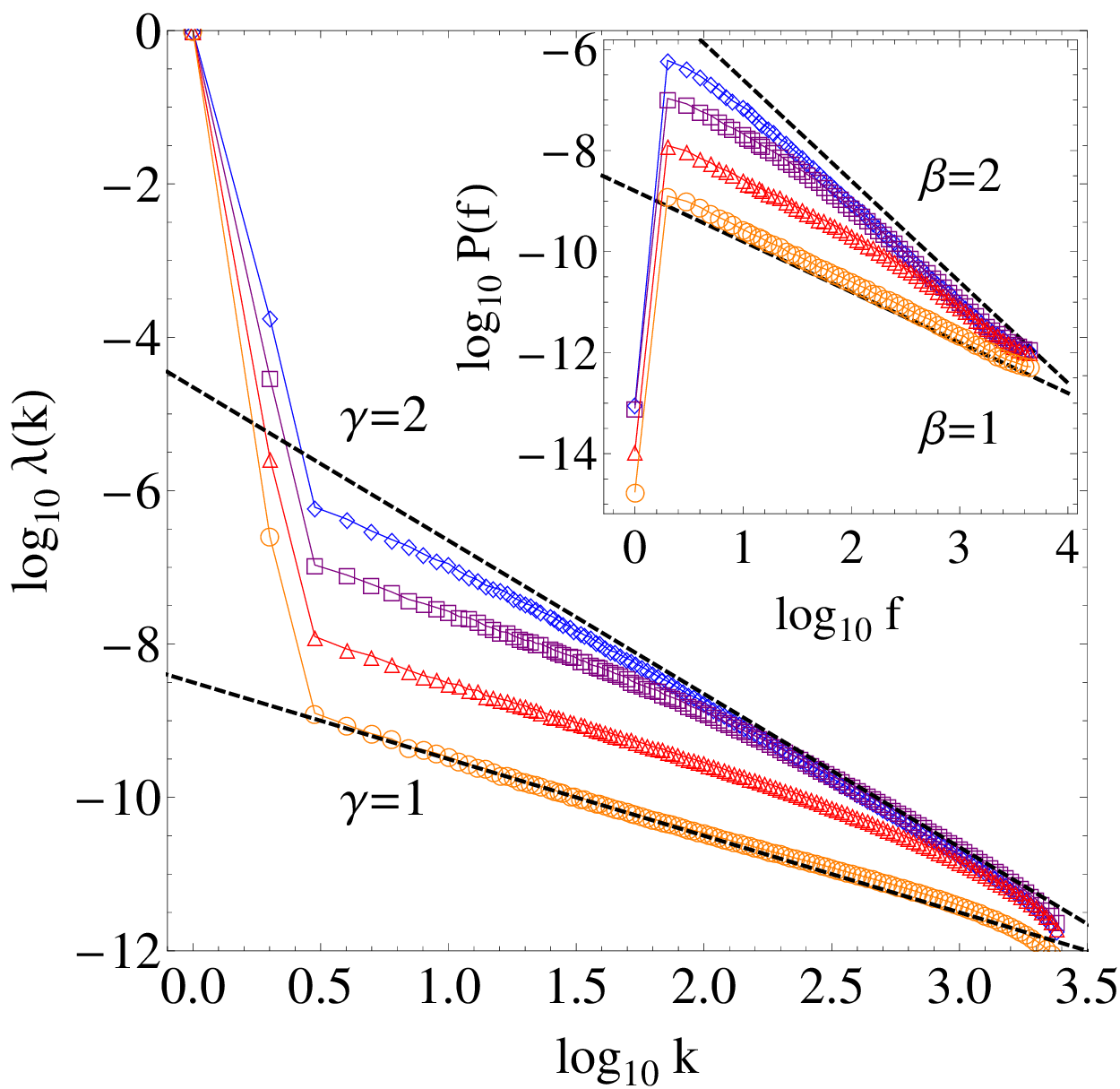}
\end{minipage}
\end{center}
\caption{Scree diagrams of ordered partial variances $\lambda_k$ for $\nu-$Hermite ensembles of $M=2500$ spectra with $N=10^4$ levels each, for $\nu=1.0$ (circles), $\nu=0.1$ (triangles), $\nu=0.01$ (squares) and $\nu=0.001$ (diamonds). Using the same symbols, the inset displays the Fourier power spectrum $P(f)$ of the fluctuations of the separate eigenspectra, averaged over the whole ensemble after individual unfolding is applied.}
\label{Fig1}
\end{figure}

\begin{figure}[t]
\begin{center}
\begin{minipage}{0.8\linewidth}
\centering
\includegraphics[width=\linewidth]{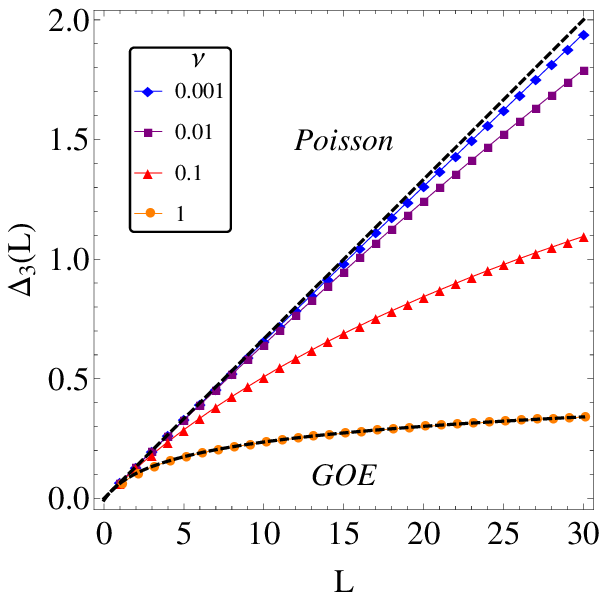}
\end{minipage}
\hfill
\begin{minipage}{0.8\linewidth}
\centering
\includegraphics[width=\linewidth]{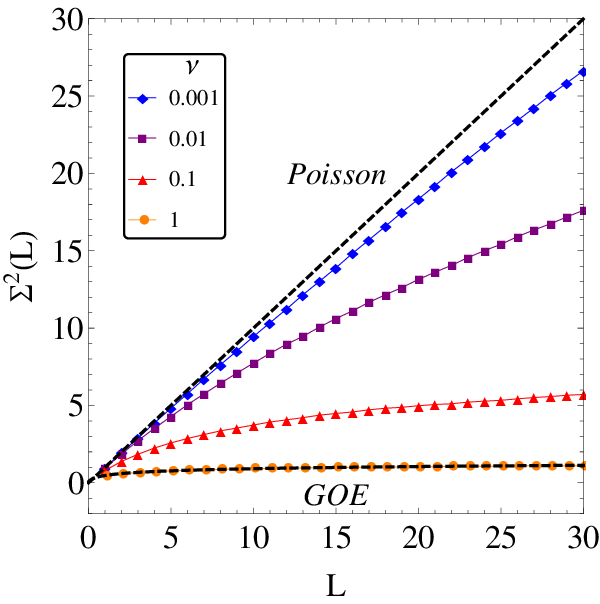}
\end{minipage}
\end{center}
\caption{$\Delta_3$ statistic (upper) and number variance $\Sigma^2$ (lower) for the $\nu-$Hermite ensembles used to obtain the scree diagrams and the Fourier power spectra shown in Fig. \ref{Fig1}. The results correspond to ensemble averages.}
\label{combo1}
\end{figure}

It is known that for finite $N$, it can be difficult to unfold the eigenspectrum analytically. Therefore, in \cite{mal07}, the unfolding was performed numerically as a polynomial fit to the ensemble-averaged level density, and then, the unfolded fluctuations were studied with Daubechies wavelets. Furthermore, it was checked that other types of wavelets lead to the similar results. In \cite{rel08} a double unfolding was performed, first unfolding with Wigner's semicircle law, and afterwards reunfolding by means of a fit with Chebyshev polynomials. After this, Fourier spectral analysis was applied to the unfolded fluctuations. Both studies confirm that the interpolation between the Poisson and GOE limits is heterogeneous, with soft behavior ($1/f^2$ power spectrum) at the finest scales, and rigid behavior ($1/f$ power spectrum) at the coarsest scales. In the analytical calculations, the cross-over frequency is predicted to occur at $f_\mathrm{\times}=\nu N/2$, whereas in the numerical calculations the cross-over is smoothed \cite{rel08}.
\begin{figure}[t]
\begin{minipage}[t]{0.45\linewidth}
\centering
  \includegraphics[width=\linewidth]{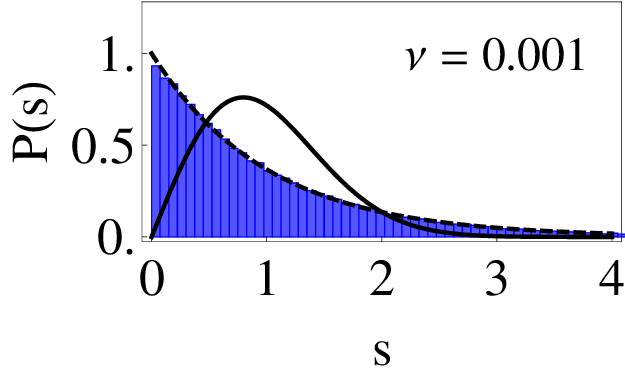}
\end{minipage}
\hfill
\begin{minipage}[t]{0.435\linewidth}
\centering
  \includegraphics[width=\linewidth]{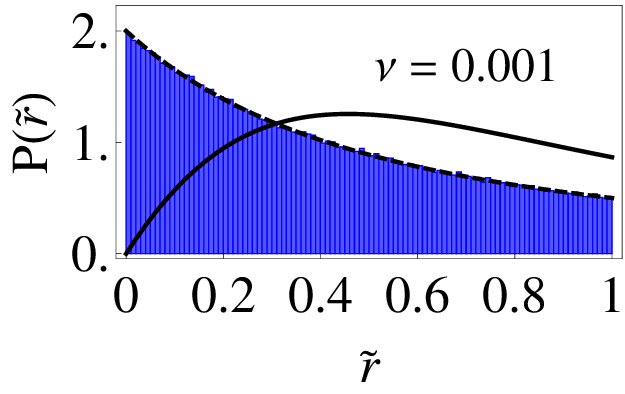}
\end{minipage}
\begin{minipage}[t]{0.45\linewidth}
\centering
  \includegraphics[width=\linewidth]{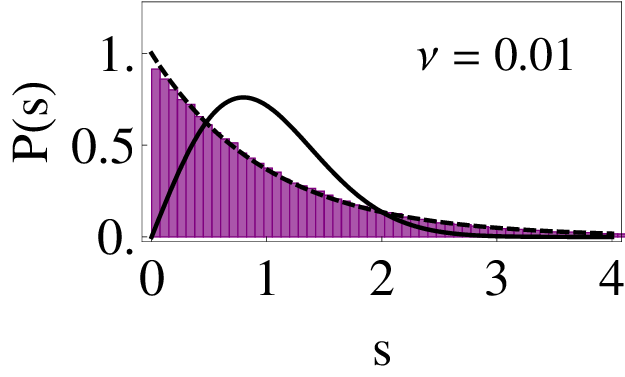}
\end{minipage}
\hfill
\begin{minipage}[t]{0.435\linewidth}
\centering
  \includegraphics[width=\linewidth]{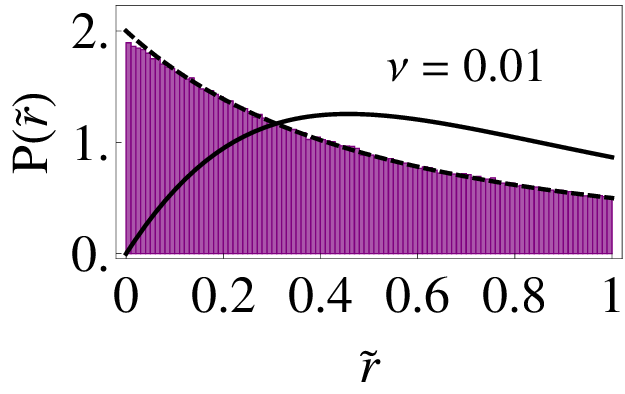}
\end{minipage}
\begin{minipage}[t]{0.45\linewidth}
\centering
  \includegraphics[width=\linewidth]{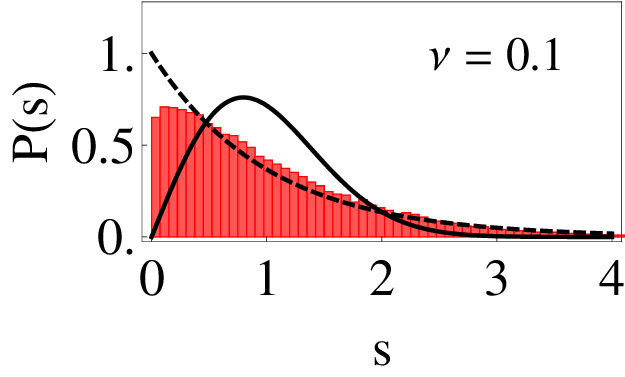}
\end{minipage}
\hfill
\begin{minipage}[t]{0.435\linewidth}
\centering
  \includegraphics[width=\linewidth]{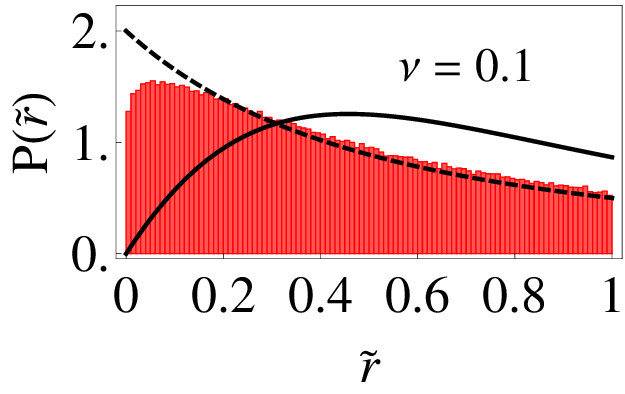}
\end{minipage}
\begin{minipage}[t]{0.45\linewidth}
\centering
  \includegraphics[width=\linewidth]{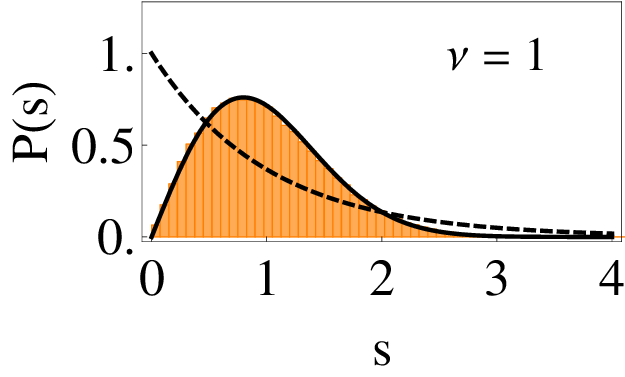}
\end{minipage}
\hfill
\begin{minipage}[t]{0.435\linewidth}
\centering
  \includegraphics[width=\linewidth]{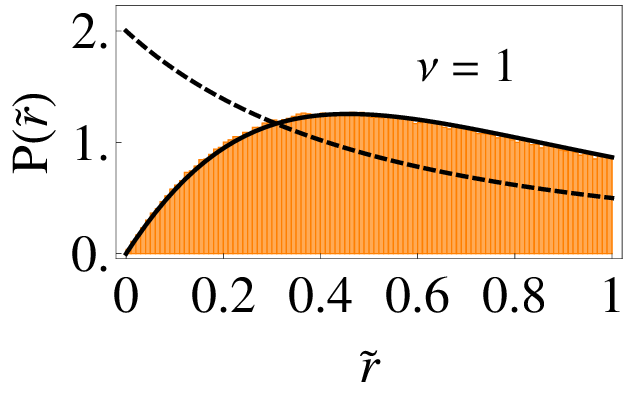}
\end{minipage}
\caption{(left) Nearest neighbor spacing distribution (NNSD) and (right) distribution $P(\tilde{r})$ for the $\nu-$Hermite ensambles used to obtain the $\Delta_3$ statistics and number variance $\Sigma^2$ shown in Fig. \ref{combo1}.}
\label{combo2}
\end{figure}

In Fig.~\ref{Fig1}, it is shown the scree diagram obtained applying SVD to a $\nu-$Hermite ensemble of $M=2500$ spectra. As the cross-over becomes more obvious for large dimensions \cite{mal07,rel08}, spectra with $N=10^4$ levels are chosen. As we can see, the scree diagram clearly distinguishes between the first $k=1,2$ trend components and the $k=3,\ldots,r$ higher-order fluctuation components, with $r=2500$. The fluctuation part of the scree diagram shows the aforementioned cross-over behavior, which is confirmed by the Fourier power spectrum of the fluctuations of the individual eigenspectra (the lowest frequency is the constant DC term). In the present approach, possible artifacts introduced by applying traditional unfolding techniques are avoided, e.g., the several points in the Fourier spectrum at low frequencies that fall far below the theoretical predictions in Ref.~\cite{rel08}. The normal modes are generated by the data themselves, avoiding the necessity to compare between different models, as in the case of studies with user-defined wavelets as in Ref.~\cite{mal07}.

In order to obtain results for the traditional fluctuation measures, we perform the data-adaptive unfolding of the spectra $E^{(m)}(n)$ using the global part $\overline{E}(n)$ calculated when we applied SVD to the ensemble \cite{fos13}. In Fig. \ref{combo1}, we show the results obtained for the long-range fluctuation measures, $\Delta_3$ statistics and the number variance $\Sigma^2$. In Fig. \ref{combo2} we compare the results of calculating the short-range fluctuation measure, NNSD, with the the distribution of the ratio of two consecutive level spacings $P(\tilde{r})$ (which, as same as the scree diagram, not require unfolding to calculate it) \cite{oga07,ata13}. Here, $\tilde{r}$ is defined as
\begin{equation}
\tilde{r}_n=\frac{\min(s_n,s_{n-1})}{\max(s_n,s_{n-1})}
\end{equation}
where $s_n=E(n+1)-E(n)$. As we can see, the results for NNSD, obtained after applying the data-adaptive unfolding, characterize the transition between the extreme regular and chaotic cases in a very similar way to that obtained with the distribution $P(\tilde{r})$.

\section{Sparse matrix ensemble}
\label{Sect_Sparse}

\begin{figure}[b]
\begin{center}
\begin{minipage}{0.9\linewidth}
\centering
\includegraphics[width=\linewidth]{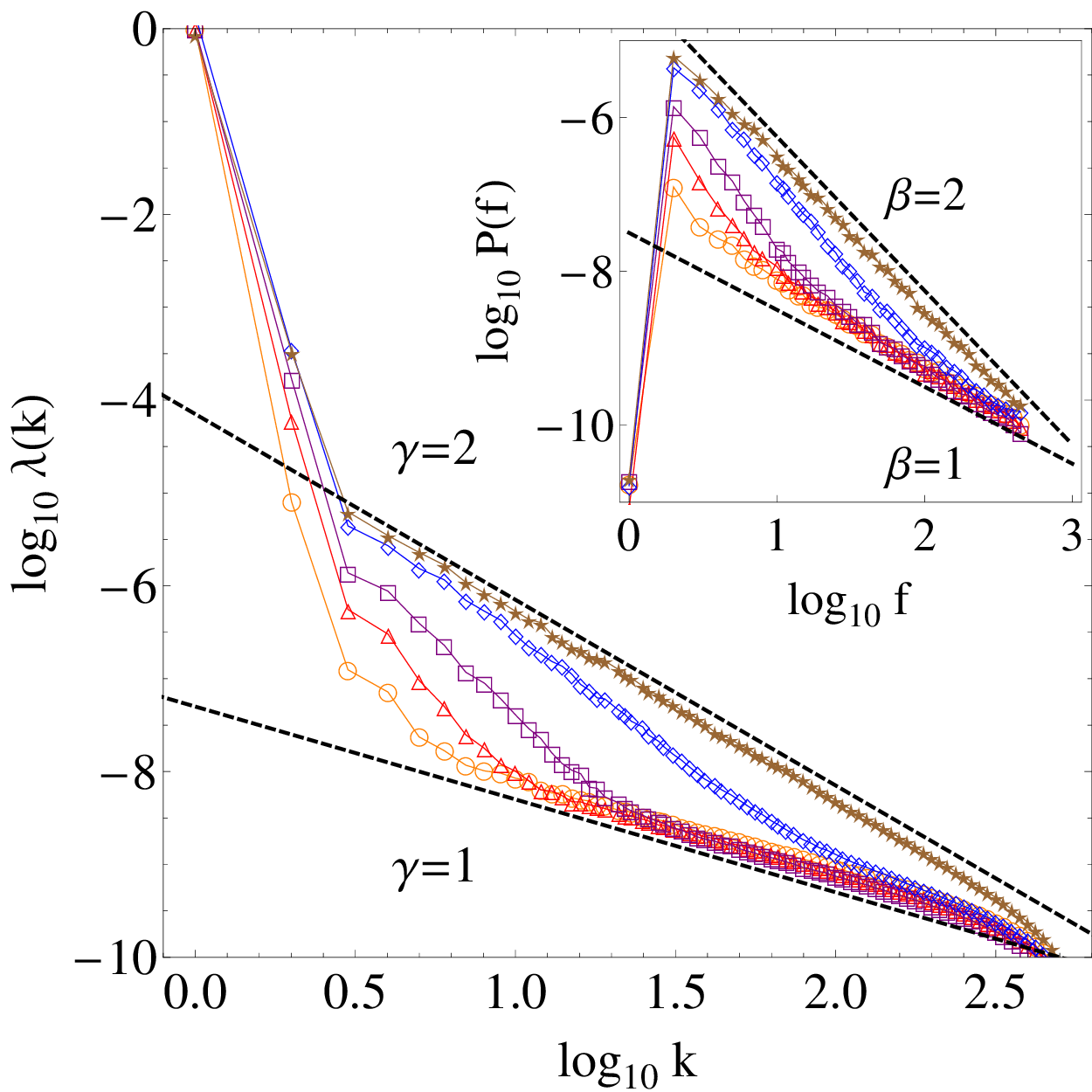}
\end{minipage}
\end{center}
\caption{Similar to Fig. \ref{Fig1}, but for ensembles of $M=500$ sparse matrices of dimension $N=2000$, for $s=0.025$ (circles), $s=0.01$ (triangles), $s=0.005$ (squares), $0.0025$ ( diamonds) and $s=0.0005$ (asterisks). In both figures, the total variance of the ensemble has been rescaled to unit variance $\sum_k \lambda_k=1$ to allow comparison between ensembles with different $s$ or $\nu$.}
\label{Fig2}
\end{figure}

\begin{figure}[t]
\begin{center}
\begin{minipage}{0.8\linewidth}
\centering
\includegraphics[width=\linewidth]{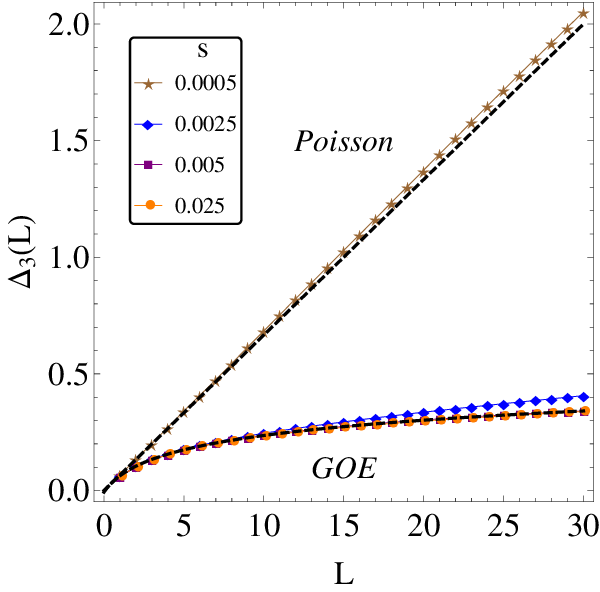}
\end{minipage}
\hfill
\begin{minipage}{0.8\linewidth}
\centering
\includegraphics[width=\linewidth]{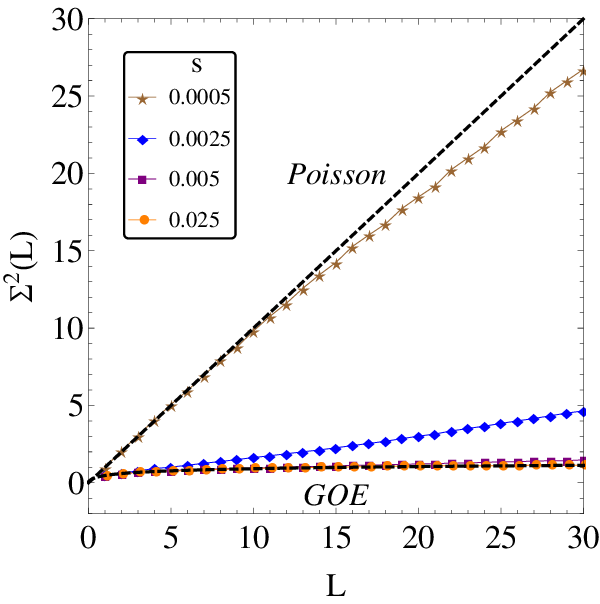}
\end{minipage}
\end{center}
\caption{$\Delta_3$ statistic (upper) and number variance $\Sigma^2$ (lower) for the sparse-matrix ensemble used to obtain the scree diagrams and the Fourier power spectra shown in Fig. \ref{Fig2}. The results correspond to ensemble averages.}
\label{combo3}
\end{figure}

In Ref.~\cite{jac01}, an ensemble of sparse real symmetric matrices was proposed starting from GOE, using a sparsity parameter $s$ which is the fraction of the $N(N-1)/2$ independent off-diagonal matrix elements chosen to be non-vanishing. All diagonal elements are kept nonzero. The non-vanishing matrix elements $H_{ij}$ are chosen independently and at random from a Gaussian distribution,
\begin{equation}
P(H_{ij})=\frac{1}{\sqrt{2 \pi \sigma_{ij}^2}} \exp \left( - \frac{H_{ij}^2}{2 \sigma_{ij}^2} \right),
\label{EqSparsityMatrix}
\end{equation}
with $\sigma_{ij}=1+\delta_{ij}$. GOE statistics is recovered in the limiting case of null sparsity $s=1$ and Poisson statistics in the case of maximum sparsity $s=0$. For arbitrary sparsity, the spectral density $\rho(E)$ is intermediate between a Gaussian shape and a semicircle, and is difficult to describe analytically. Thus, Ref.~\cite{jac01} carried out a numerical unfolding, fitting a polynomial of arbitrary degree to the integrated level density of either a single realization of the eigenspectrum (individual-spectrum unfolding or self unfolding), or to the integrated level density averaged over the whole ensemble (ensemble unfolding). After this prior unfolding, a normal-mode analysis similar to Eqs.~(\ref{EqX})-(\ref{EqSVD}) was applied to the ensemble of fluctuations, in which case only fluctuation normal modes and no trend normal modes are obtained. For intermediate sparsities $0<s<1$, instead of the scale invariance property of Eq.~(\ref{EqScree}), a cross-over was observed for the fluctuations with rigid behavior ($\gamma=1$) for higher order-numbers $k$ (finer scales) and soft behavior ($\gamma=2$) at lower order-numbers (coarser scales). The location of the cross-over $k_\times$ was found to be proportional with the dimension of the spectrum, $k_\times \propto \sqrt{N}$, and shifts to higher order-numbers $k$ (finer scales) for lower sparsities $s$. An ambiguity was found in the number variance fluctuation measure $\Sigma^2$ when calculated after individual-spectrum unfolding or after ensemble unfolding, which however was argued to be an artificial effect of the two types of unfolding that were compared, and Ref~\cite{jac01} warned against interpreting erroneously the sparse matrix model of being non-ergodic.

\begin{figure}[t]
\begin{minipage}[t]{0.45\linewidth}
\centering
  \includegraphics[width=\linewidth]{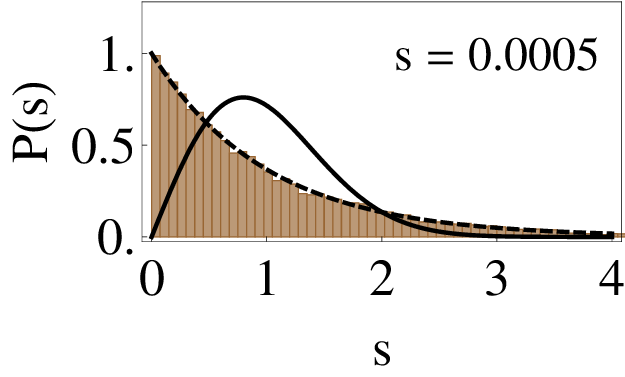}
\end{minipage}
\hfill
\begin{minipage}[t]{0.43\linewidth}
\centering
  \includegraphics[width=\linewidth]{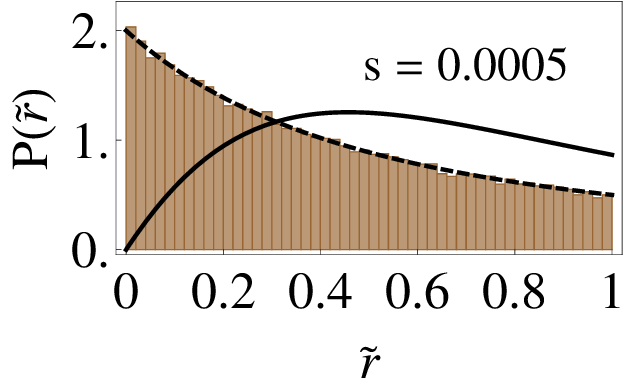}
\end{minipage}
\begin{minipage}[t]{0.45\linewidth}
\centering
  \includegraphics[width=\linewidth]{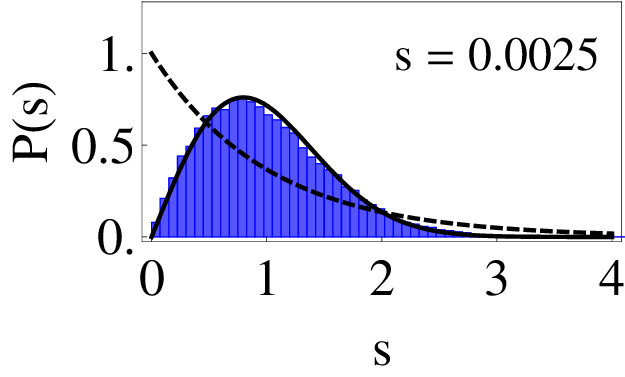}
\end{minipage}
\hfill
\begin{minipage}[t]{0.43\linewidth}
\centering
  \includegraphics[width=\linewidth]{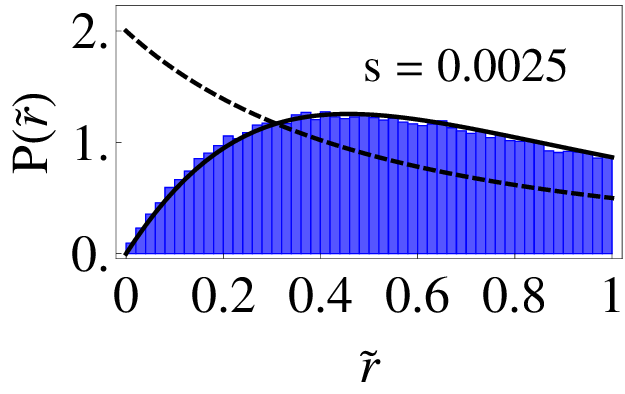}
\end{minipage}
\begin{minipage}[t]{0.45\linewidth}
\centering
  \includegraphics[width=\linewidth]{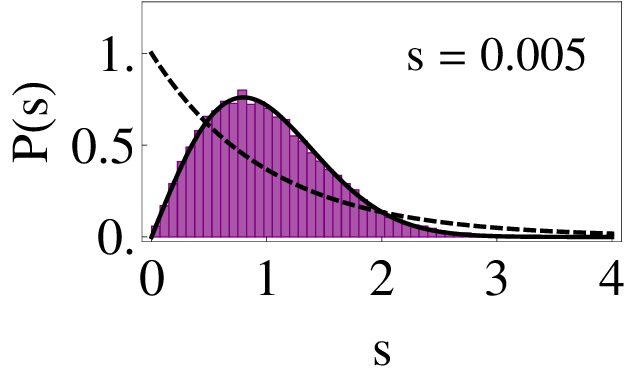}
\end{minipage}
\hfill
\begin{minipage}[t]{0.43\linewidth}
\centering
  \includegraphics[width=\linewidth]{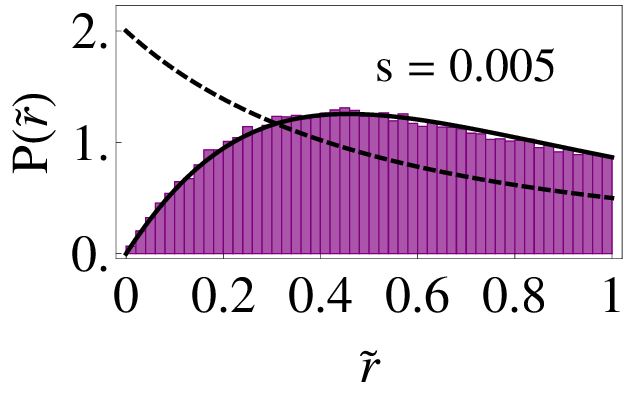}
\end{minipage}
\begin{minipage}[t]{0.45\linewidth}
\centering
  \includegraphics[width=\linewidth]{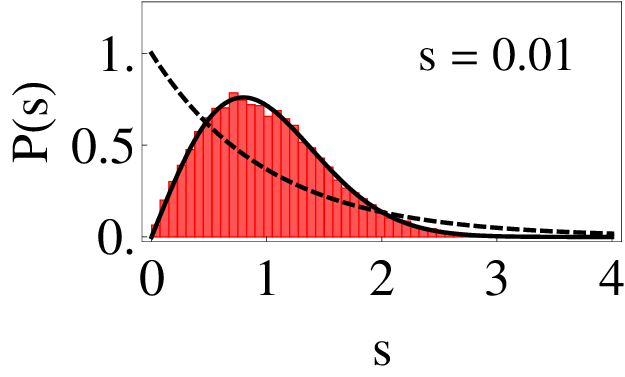}
\end{minipage}
\hfill
\begin{minipage}[t]{0.43\linewidth}
\centering
  \includegraphics[width=\linewidth]{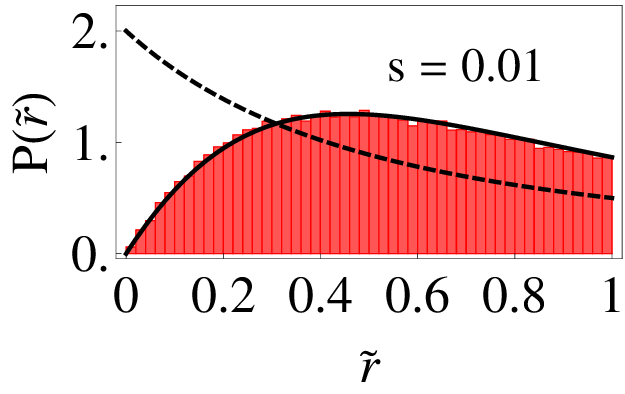}
\end{minipage}
\begin{minipage}[t]{0.45\linewidth}
\centering
  \includegraphics[width=\linewidth]{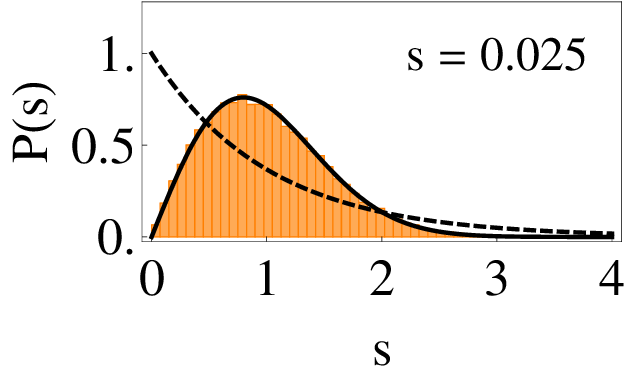}
\end{minipage}
\hfill
\begin{minipage}[t]{0.43\linewidth}
\centering
  \includegraphics[width=\linewidth]{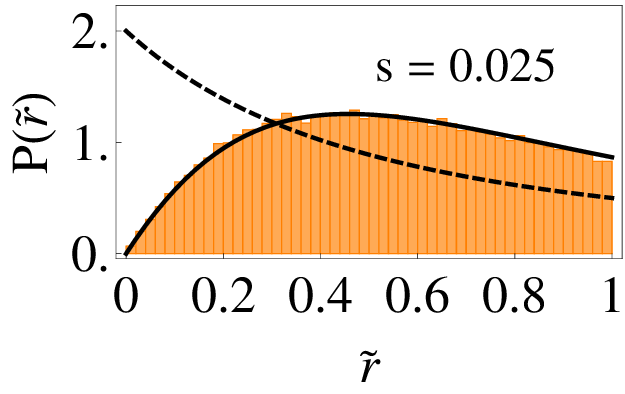}
\end{minipage}
\caption{(left) Nearest neighbor spacing distribution (NNSD) and (right) distribution $P(\tilde{r})$ for the sparse-matrix ensambles used to obtain the $\Delta_3$ statistics and number variance $\Sigma^2$ shown in Fig. \ref{combo3}.}
\label{combo4}
\end{figure}

In Fig.~\ref{Fig2}, we show the results of applying SVD to an ensemble of $M=500$ spectra of sparse matrices, each spectrum containing $N=2000$ levels. The optimal ensemble size is $M \approx N/4$, because for $M \gg N/4$ there is a long tail of insignificant partial variances in the scree diagram, whereas for $M \ll N/4$ the range of scales of the scree diagram becomes restrained, but the statistical properties do not depend on the particular choices of $N$ and $M$ \cite{fos13}. The scree diagram clearly distinguishes between the first $k=1,2$ trend modes and $k=3,\ldots,r$ higher-order fluctuations modes, with $r=500$. The fluctuation part of the scree diagram exhibits the cross-over behavior, and $k_\mathrm{\times}$ shifts towards higher order-numbers (finer scales) for decreasing sparsity, in correspondence to Ref.~\cite{jac01}. On the other hand, after individual unfolding of the separate spectra of the ensemble, according to Eq.~(\ref{EqDecomposition}), the same cross-over behavior is observed in the Fourier power spectrum (note again that the lowest frequency is the constant DC term and does not belong to the fluctuations). The scree diagram is an ensemble-averaged property and shows the collective behavior of the normal modes common to the whole ensemble, whereas the Fourier power spectrum of the individual fluctuations is an individual-spectrum averaged property. Thus, in the present framework, ensemble-averaged and individual-spectrum averaged statistics can be studied consistently within the same basis of normal modes, and avoid the ambiguities of the standard unfolding applied in Ref.~\cite{jac01}.

Finally, as same as for the $\nu-$Hermite ensemble, in order to calculate the traditional fluctuation measures we performed the data-adaptive unfolding of the spectra. In Fig. \ref{combo3}, we show the results obtained for $\Delta_3$ statistics and the number variance $\Sigma^2$. In Fig. \ref{combo4} we show the results for NNSD and $P(\tilde{r})$. As before, the results for NNSD, characterize the transition from GOE to Poisson statistics in a very similar way to that obtained with the distribution $P(\tilde{r})$, which does not require an unfolding procedure to calculate it.

\section{Conclusions}

We have applied directly SVD to non-standard random-matrix spectra of the $\nu-$Hermite and sparse matrix ensembles. The reported cross-over of the fluctuation statistics is obtained without performing previously any unfolding procedure, avoiding the introduction of possible artifacts. We also confirm again that the data-adaptive unfolding, implemented employing SVD, works well even for non-standard random-matrix spectra, in such a way the traditional spectral fluctuation measures can be systematically calculated. Moreover, the application of SVD allowed us to calculate and compare ensemble-averaged and individual-spectrum averaged statistics in a consistent way within the same basis of normal modes.

\section*{Acknowledgements}

We acknowledge financial support from CONACyT (Grant No. CB-2011-01-167441). G.T.V. acknowledges CONACyT for support through a scholar fellowship (Grant No. 381047).

\end{document}